\documentclass[12pt]{article}

\renewcommand{\d}{{\rm d}}

\newcommand{\son}{\sigma^1}
\newcommand{\tson}{\tilde{\sigma}^1}

\begin{document}

\title{Faddeev formulation of gravity in discrete form}
\author{V.M. Khatsymovsky \\
 {\em Budker Institute of Nuclear Physics} \\ {\em
 Novosibirsk,
 630090,
 Russia}
\\ {\em E-mail address: khatsym@inp.nsk.su}}
\date{}
\maketitle
\begin{abstract}
We study Faddeev formulation of gravity, in which the metric is composed of vector fields. We consider these fields constant in the interior of the 4-simplices of a simplicial complex. The action depends not only on the values of the fields in the interior of the 4-simplices but on the details of (regularized) jump of the fields between the 4-simplices. Though, when the fields vary arbitrarily slowly from the 4-simplex to 4-simplex, the latter dependence is negligible (of the next-to-leading order of magnitude).

We put the earlier proposed in our work first order (connection) representation of the Faddeev action into the discrete form. We show that upon excluding the connections it is consistent with the above Faddeev action on the piecewise constant fields in the leading order of magnitude. Thus, using the discrete form of the connection representation of the Faddeev action can serve a way to fix the value of this action on the piecewise constant ansatz on simplices.
\end{abstract}

keywords: Einstein theory of gravity; connection; piecewise flat spacetime


MSC classes: 83C27; 53C05







\section{ Introduction}

Recently Faddeev has proposed \cite{Fad} a new formulation of Einstein's gravity described by the set of ten covariant vector fields $f^A_\lambda (x)$, $A = 1, \dots , 10$. The metric is a composite field, $g_{\lambda \mu} = f^A_\lambda f_{\mu A}$. The action takes the form
\begin{equation}\label{Fad action}                                          
S = \int {\cal L} \d^4 x = \int \Pi^{AB} (f^\lambda_{A, \lambda} f^\mu_{B, \mu} - f^\lambda_{A, \mu} f^\mu_{B, \lambda}) \sqrt {g} \d^4 x.
\end{equation}

\noindent Here $\Pi_{AB} = \delta_{AB} - f^\lambda_A f_{\lambda B}$, raising and lowering world (greek) indices is performed with the help of $g_{\lambda \mu}$, and $f^\lambda_{A, \mu} \equiv \partial_\mu f^\lambda_A$.

Faddeev action for gravity has the following properties which may provide certain advantage of using this formulation instead of the standard metric one based on the Hilbert-Einstein action. First, suppose metric has a discontinuity along the coordinate $x^n$ such that this discontinuity is experienced by the components of metric transverse to the direction of $x^n$: $g_{t_1 t_2 , n} \sim \delta (x^n)$ is $\delta$-function like. The Hilbert-Einstein action is divergent: ${\cal L}$ contains contribution $\sim (g_{t_1 t_2 , n})^2 \sim (\delta (x^n))^2$, $\int {\cal L} \d^4 x = \infty$. The Faddeev action is not divergent in this case since there is no the square of any derivative in the action. Second, suppose vertical components of the classical equations of motion are fulfilled and are non-degenerate leading to
\begin{equation}\label{T=0}                                                 
T_{\lambda , [\mu \nu ]} = f^A_\lambda (f_{\mu A , \nu} - f_{\nu A , \mu}) = 0.
\end{equation}

\noindent Consider the plane dividing a region of the spacetime into the two regions each with constant metric. The coordinate dependence of the metric in this region (stepwise) is only on the coordinate $x^n$ normal to the plane. This fact together with equation (\ref{T=0}) means vanishing discontinuity of the metric,
\begin{equation}                                                            
g_{t_1 t_2 , n} = f_{t_2}^A f_{t_1 A , n} + f_{t_1}^A f_{t_2 A , n} = f_{t_2}^A f_{n A , t_1} + f_{t_1}^A f_{n A , t_2} + T_{t_2 , [t_1 n]} + T_{t_1 , [t_2 n]} = 0.
\end{equation}

\noindent (According to the above assumption in this example, the derivatives over the coordinates $x^{t_1}, x^{t_2}$ lying in the plain are zero). Thus, the discontinuity of the transverse components of metric is allowed in the Faddeev gravity, but only virtually, on quantum level. The situation is illustrated by the table \ref{discontinuities}. Important simplifying consequence for description of quantum Faddeev gravity by piecewise-constant fields ($f_{\lambda A}$) is that we do not need to impose conditions requiring continuity of metric induced on the face shared by any two regions in each of which the fields are constant. In other words, these regions a priori may not coincide on their common face, and the values $f_{\lambda A}$ may be chosen freely in each region of their constancy. In turn, piecewise-constant distribution of the fields $f_{\lambda A}$ seems to be an appropriate ansatz for studying the "gas" of metric discontinuities.

\begin{table}
\caption{Possibility of discontinuity of the transverse components of metric.}
\label{discontinuities}
\begin{tabular}{|l|c|c|}          \hline
~~~~~~~~~~formulation & Hilbert      & Faddeev            \\
framework  & -Einstein     &  \\ \hline
classical &  ---   & --- \\
quantum  & --- & + \\
\hline
          \end{tabular}
\unitlength 1pt
\begin{picture}(0,0)
\thicklines
\put(-226,46){\line(5,-2){107}}
\end{picture}
\end{table}

\begin{figure}
\unitlength 1pt
\begin{picture}(150,150)(-100,20)
\put(30,30){\line(1,0){40}}
\put(30,30){\line(0,1){40}}
\put(30,70){\line(1,0){40}}
\put(30,70){\line(1,4){10}}

\put(40,110){\vector(1,4){5}}
\put(30,125){$x^2$}

\put(70,30){\line(4,-1){40}}

\put(110,20){\vector(4,1){20}}
\put(130,18){$x^1$}

\put(70,30){\line(0,1){40}}
\put(70,70){\line(4,-1){40}}

\put(70,70){\line(1,4){10}}
\put(40,110){\line(1,0){40}}
\put(80,110){\line(4,-1){40}}

\put(110,60){\line(1,4){10}}
\put(110,20){\line(0,1){40}}

\put(42,60){$(1,1)$}
\put(3,60){$(0,1)$}
\put(82,50){$(2,1)$}
\put(95,10){$(2,0)$}
\put(50,20){$(1,0)$}
\put(6,20){$(0,0)$}
\put(92,90){$(2,2)$}
\put(50,99){$(1,2)$}
\put(12,102){$(0,2)$}

\end{picture}
\caption{The manifold composed of polyhedrons with the topology of a cube.}
\label{cubes}
\end{figure}
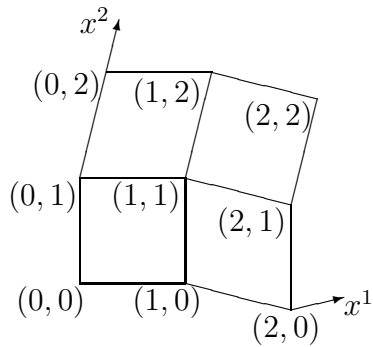

As is known, piecewise flat Riemannian manifold can be represented as simplicial complex where the metric can be chosen constant in each simplex \cite{piecewise flat=simplicial}. If one considers the manifold composed of polyhedrons with the topology of a cube (fig. \ref{cubes}),
requirements that the metric be constant in each cube and transverse components of metric be continuous on each cubic face lead to essential restriction on possible form of metric. Indeed, introduce piecewise affine world coordinates $x^\lambda$ such that vertices have integer coordinates and any cube edge is described by one of the four vectors $(1, 0, 0, 0)$, $(0, 1, 0, 0)$, $(0, 0, 1, 0)$, $(0, 0, 0, 1)$. Passing through the chain of the cubes along the coordinate, say, $x^1$, we find that $g_{\lambda \mu}$ cannot depend on $x^1$ at $\lambda \neq 1$ and $\mu \neq 1$. Analogous conclusions can be made for other coordinates, and we can write the coordinate dependence of metric tensor components as $g_{11} (x^1 )$, $g_{12} (x^1, x^2)$, \dots . Of course, this form of metric in no way can be regarded as general one, and this metric does not provide proper ansatz for minisuperspace gravity system. This corresponds to intuitive feeling that we cannot approximate general curved manifold by flat polyhedrons with the topology of a cube.

However, situation becomes qualitatively different if discontinuities of the transverse components of metric are allowed as in the considered more general case of the piecewise constant $f_{\lambda A}$ in the Faddeev gravity. In this case the above speculation does not give any restriction on the form of metric which can be approximated by the cube-like polyhedrons, and the ansatz based on cubic decomposition of spacetime may be of interest. This ansatz is considerably more simple than the simplicial one, and remind the usual lattice discretization.

\section{Faddeev action on piecewise constant fields}

Let us try to write out the Faddeev action (\ref{Fad action}) for the piecewise-constant fields on simplicial complex. Let $x^\lambda$ be piecewise-affine coordinate frame; $f^\lambda_A (x ) = const$ in the interior of every 4-simplex $\sigma^4$. The field $f^\lambda_A$ in the most part of the neighborhood of any 3-simplex depends (in stepwise manner) only on one (normal to $\sigma^3$) coordinate. Therefore contribution to $S$ from $\sigma^3$ is zero. Contributions to $S$ come from the neighborhood of the 2-simplices $\sigma^2$ due to a dependence on the two coordinates, say, $x^1$, $x^2$. Evidently, the expression $(f^\lambda_{A, \lambda} f^\mu_{B, \mu} - f^\lambda_{A, \mu} f^\mu_{B, \lambda})$ appearing in $S$ has support on $\sigma^2$. That is, it is $\delta$-function $const \cdot \delta (x^1) \delta (x^2)$. The constant can be reliably defined with taking into account the fact that this expression is the full derivative,
\begin{equation}                                                            
f^\lambda_{A, \lambda} f^\mu_{B, \mu} - f^\lambda_{A, \mu} f^\mu_{B, \lambda} = \partial_\lambda Q^\lambda, ~~~ Q^\lambda = f^\lambda_A \partial_\mu f^\mu_B - f^\mu_B \partial_\mu f^\lambda_A.
\end{equation}

\noindent Then integral over any neighborhood of the point $(x^1, x^2) = (0, 0)$ (which defines this constant) reduces to the contour integral not depending on the details of the behavior of the fields at this point. Taking into account the subsequent symmetrization over $A, B$ we have
\begin{equation}\label{oint_C}                                              
\int (f^\lambda_{A, \lambda} f^\mu_{B, \mu} - f^\lambda_{A, \mu} f^\mu_{B, \lambda}) \d x^1 \d x^2 = \oint_C (f^1_A \d f^2_B - f^2_B \d  f^1_A).
\end{equation}

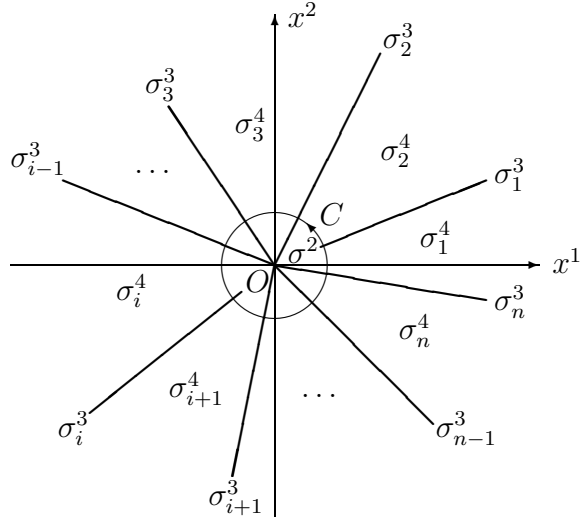
\begin{figure}
\unitlength 1pt
\begin{picture}(200,200)(-200,-100)
\put(0,0){\line(-1,0){100}}
\put(0,0){\line(0,-1){95}}
\put(0,0){\vector(1,0){100}}
\put(0,0){\vector(0,1){95}}

\put(0,0){\circle{40}}
\put(16,12){\vector(-1,1){4}}
\put(83,31){$\sigma^3_1$}
\put(41,82){$\sigma^3_2$}
\put(5,90){$x^2$}
\put(105,-3){$x^1$}
\put(-49,65){$\sigma^3_3$}
\put(-100,37){$\sigma^3_{i-1}$}
\put(-82,-65){$\sigma^3_i$}
\put(-25,-91){$\sigma^3_{i+1}$}
\put(61,-65){$\sigma^3_{n-1}$}
\put(83,-17){$\sigma^3_n$}
\put(5,1){$\sigma^2$}
\put(-11,-10){$O$}
\put(17,15){$C$}
\put(40,40){$\sigma^4_2$}
\put(55,7){$\sigma^4_1$}
\put(-15,50){$\sigma^4_3$}
\put(-53,35){\dots}
\put(-60,-13){$\sigma^4_i$}
\put(-40,-50){$\sigma^4_{i+1}$}
\put(10,-50){\dots}
\put(47,-28){$\sigma^4_n$}

\thicklines

\put(17.5,7){\line(5,2){62.5}}
\put(0,0){\line(1,2){40}}
\put(0,0){\line(-2,3){40}}
\put(0,0){\line(-5,2){80}}
\put(-12.5,-10){\line(-5,-4){57.5}}
\put(0,0){\line(-1,-5){16}}
\put(0,0){\line(6,-1){80}}
\put(0,0){\line(1,-1){60}}

\end{picture}

\caption{The neighborhood of a triangle $\sigma^2$ shared by 3- and 4-simplices.}
\label{sigma2}
\end{figure}

\noindent In fig. \ref{sigma2} the center $O$ which represents the 2-simplex $\sigma^2$ is encircled by the integration contour $C$ counterclockwise.

There are products of the step functions and delta functions under the contour integral sign in (\ref{oint_C}) which can be defined ambiguously depending on the intermediate regularization. Formally, we can write $\theta (x) \delta (x) = \theta (0) \delta (x)$ where we can take for $\theta (0)$ any number $\alpha$ from the interval $[0,1]$. In the geometry of fig. \ref{sigma2} this amounts to the choice of the value of a function on the 3-face $\sigma^3_i$ on which it undergoes the discontinuity when passing between the two 4-simplices $\sigma^4_i$ and $\sigma^4_{i+1}$ sharing this 3-face,
\begin{equation}                                                            
f(\sigma^3_i) = (1 - \alpha ) f(\sigma^4_i ) + \alpha f(\sigma^4_{i+1} ).
\end{equation}

\noindent Here $f$ is $f^1_A$ or $f^2_B$. As a result, the value of the integral (\ref{oint_C}) is
\begin{eqnarray}\label{delta f delta f}                                     
& & \sum^n_{i = 1} \left \{ [ \alpha f^1_A (\sigma^4_{i+1} ) + (1 - \alpha ) f^1_A (\sigma^4_i ) ] [ f^2_B (\sigma^4_{i+1} ) - f^2_B (\sigma^4_i ) ] \right. \phantom{ f^2_B (\sigma^4_i ) ] [ f^1_A (\sigma^4_{i+1} ) - f^1_A (\sigma^4_i ) ]} \nonumber \\ & & \phantom{[ \alpha f^1_A (\sigma^4_{i+1} ) + (1 - \alpha ) } \left. - [ \alpha f^2_B (\sigma^4_{i+1} ) + (1 - \alpha ) f^2_B (\sigma^4_i ) ] [ f^1_A (\sigma^4_{i+1} ) - f^1_A (\sigma^4_i ) ] \right \} \nonumber \\ & & \hspace{10mm} = \sum^n_{i=1} [ f^1_A ( \sigma^4_i ) f^2_B ( \sigma^4_{i+1} ) - f^1_A ( \sigma^4_{i+1} ) f^2_B ( \sigma^4_i ) ].
\end{eqnarray}

\noindent Remarkable is that the dependence on $\alpha$ disappears. Thus,
\begin{equation}\label{d f d f}                                             
f^\lambda_{A, \lambda} f^\mu_{B, \mu} - f^\lambda_{A, \mu} f^\mu_{B, \lambda} = \delta (x^1 ) \delta (x^2 ) \sum^n_{i=1} [ f^1_A ( \sigma^4_i ) f^2_B ( \sigma^4_{i+1} ) - f^1_A ( \sigma^4_{i+1} ) f^2_B ( \sigma^4_i ) ].
\end{equation}

\noindent The symmetrization over $A, B$ is implied.

Let $\delta f$ be typical variation of $f^\lambda_A$ when passing from simplex to simplex. Note that eq. (\ref{delta f delta f}) has the order of magnitude $O((\delta f)^2)$. If there is certain fixed continuum (smooth) distribution of $f^\lambda_A$ on the fixed smooth manifold, and the considered piecewise flat geometry is only approximation to this continuum one which is becoming more and more fine, this order of magnitude $O((\delta f)^2)$ of (\ref{delta f delta f}) just should reproduce the continuum value of action while next-to-leading orders $o((\delta f)^2)$ tend to zero.

Next we would like to multiply eq. (\ref{d f d f}) by $\Pi^{AB} \sqrt{g}$. Since the latter function is discontinuous at $(x^1, x^2) \to (0, 0)$, this product can not be defined unambiguously. We can only write
\begin{eqnarray}\label{d f d f V Pi}                                        
\hspace{-8mm} (f^\lambda_{A, \lambda} f^\mu_{B, \mu} - f^\lambda_{A, \mu} f^\mu_{B, \lambda}) \Pi^{AB} \sqrt{g} & = & \delta (x^1 ) \delta (x^2 ) \left \{\Pi^{AB} (\sigma^2 ) \sqrt{g (\sigma^2 )} \sum^n_{i=1} \left [ f^1_A ( \sigma^4_i ) f^2_B ( \sigma^4_{i+1} ) \right. \right. \nonumber \\ & & \left. \left. - f^1_A ( \sigma^4_{i+1} ) f^2_B ( \sigma^4_i ) \right ] + O((\delta f)^3) \right \}.
\end{eqnarray}

\noindent Here $\Pi^{AB} (\sigma^2 ) \sqrt{g (\sigma^2 )}$ means the value of $\Pi^{AB} \sqrt{g}$ in any one of the 4-simplices $\sigma^4_i$ sharing $\sigma^2$ (or some average of these), and $O((\delta f)^3)$ is a contribution dependent on the model of regularization of $f^\lambda_A (x )$ in the neighborhood of the point of discontinuity $(x^1, x^2) = (0, 0)$. As mentioned above, this contribution does not contribute to the action in the continuum limit.

\section{Discrete first order formalism for the Faddeev action}

\begin{figure}
\unitlength 1pt
\begin{picture}(50,50)(-150,0)

\put(0,50){\vector(1,0){53}}
\put(53,50){\line(1,0){47}}
\put(0,0){\vector(0,1){28}}
\put(0,28){\line(0,1){22}}
\put(100,50){\vector(0,-1){28}}
\put(100,22){\line(0,-1){22}}
\put(0,0){\vector(1,0){53}}
\put(53,0){\line(1,0){47}}

\put(0,0){\circle*{4}}
\put(100,0){\circle*{4}}
\put(0,50){\circle*{4}}
\put(100,50){\circle*{4}}

\put(-65,0){continuum}
\put(-65,-10){II order}
\put(-5,-10){1}

\put(-65,50){continuum}
\put(-65,40){I order}
\put(-5,53){2}

\put(115,0){discrete}
\put(115,-10){II order}
\put(100,-10){4}

\put(115,50){discrete}
\put(115,40){I order}
\put(100,53){3}
\end{picture}

\caption{Different formalisms and discretization.}
\label{discr}
\end{figure}
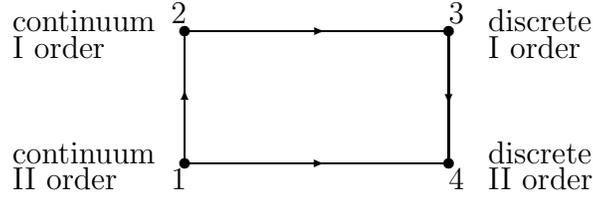

An idea of how to avoid the above model dependence of the simplicial Faddeev second order formalism is to use at an intermediate stage the first order formalism, that is, to use the connection type variables. The hope is that the action in the first order formalism contains the derivatives only linearly and, consequently, it is less singular than in the second order formalism. The diagram of fig. \ref{discr} illustrates the situation.
Above we have considered the direct discretization of the genuine second order Faddeev formalism on the piecewise constant ansatz, $1 \to 4$ in the diagram. Now we are in a position to study transition $1 \to 2 \to 3 \to 4$. Commutativity of the diagram which we shall prove below means that unambiguously defined part $O((\delta f)^2)$ of the discrete Faddeev action (sufficient to reproduce its true continuum limit) can be reproduced through intermediate use of the first order (connection) formalism. At the same time, contrary to $1 \to 4$, this transition $1 \to 2 \to 3 \to 4$ gives unambiguous result for the total discrete (simplicial minisuperspace) Faddeev action.

The first order formalism for the Faddeev formulation considered in our paper \cite{our} is the SO(10) connection representation of the Cartan-Weyl type with the additional SO(10) local symmetry violating condition on the connection $\omega_{\lambda AB}$. The action is
\begin{eqnarray}\label{S full}                                             
& & S = S_{SO(10)} + S_\omega, ~~~ S_{SO(10)} = \int f^{\lambda A} f^{\mu B} R_{\lambda \mu AB} \sqrt{g} \d^4 x, \nonumber \\ & & S = \int f^{\lambda A} f^{\mu B} ( R_{\lambda \mu AB} + \Lambda^\nu_{[\lambda \mu]} \omega_{\nu AB} ) \sqrt{g} \d^4 x, \\ & & R_{\lambda \mu AB} = \partial_\lambda \omega_{\mu AB} - \partial_\mu \omega_{\lambda AB} + (\omega_\lambda \omega_\mu - \omega_\mu \omega_\lambda)_{AB}. \nonumber
\end{eqnarray}

\noindent The $\Lambda^\nu_{[\lambda \mu]}$ are the Lagrange multipliers of the condition on $\omega_{\lambda AB}$,
\begin{equation}\label{wff}                                                
\omega_{\lambda AB} f^A_\mu f^B_\nu = 0.
\end{equation}

In the discrete Cartan-Weyl theory, the vectors of edges in the local Euclidean frames of the 4-simplices are the tetrad type variables. More accurately, the continuum analog of the edge vector is infinitesimal diffeomorphism invariant (or invariant w. r. t. the world index) $e^a_\lambda \d x^\lambda$. In the discrete theory, $e^a_\lambda$ is constant in the interior of each 4-simplex, and $\d x^\lambda$ is substituted by
a 4-vector $\Delta x^\lambda_{\sigma^1}$ in arbitrary piecewise-affine coordinates $x^\lambda$,
\begin{equation}                                                           
\Delta x^\lambda_{\sigma^1} = x^\lambda (\sigma^0_2 ) - x^\lambda (\sigma^0_1 )
\end{equation}

\noindent for the edge $\sigma^1$, the difference between the coordinates of its ending vertices $\sigma^0_1$, $\sigma^0_2$. Then the edge vector
\begin{equation}                                                           
e^a_{\sigma^1} = e^a_\lambda \Delta x^\lambda_{\sigma^1}
\end{equation}

\noindent is a value invariant w. r. t. the world index or w. r. t. the coordinates of the vertices. From this definition in the same 4-simplex (where $e^a_\lambda = const$) the closure condition (vanishing algebraic sum) for the vectors of edges of any triangle $\sigma^2$ is automatically satisfied,
\begin{equation}                                                           
\sum_{ \{ \sigma^1 : ~ \sigma^1 \subset \sigma^2 \} } \pm e^a_{\sigma^1} = 0.
\end{equation}

\noindent Thus, the covariant tetrad components are direct analogs of the variables related to geometrical elements of the simplicial complex. The usual Cartan-Weyl form of the Einstein action can be readily rewritten in terms of them as
\begin{equation}\label{Cartan}                                             
\int \epsilon^{\lambda \mu \nu \rho} \epsilon_{abcd} e^a_\lambda e^b_\mu R_{\nu \rho}{}^{cd} (\omega ) \d^4 x
\end{equation}

\noindent where $\lambda, \mu, \nu, \dots $ = 1, 2, 3, 4; ~ $a, b, c, \dots$ = 1, 2, 3, 4.

Analogously, the variables $f^A_\lambda$ with covariant world index $\lambda$ give rise to 10-di\-men\-sion\-al vectors of edges $\sigma^1$,
\begin{equation}                                                           
f^A_{\sigma^1} = f^A_\lambda \Delta x^\lambda_{\sigma^1}.
\end{equation}

\noindent The (algebraic) sum of these over the edges of any triangle in the same 4-simplex is zero,
\begin{equation}\label{Sum f}                                              
\sum_{ \{ \sigma^1 : ~ \sigma^1 \subset \sigma^2 \} } \pm f^A_{\sigma^1} = 0.
\end{equation}

\noindent Of course, the full 10-dimensional analog of $\epsilon_{abcd}$ has ten indices, but we need rather the analog in the "horizontal" 4-dimensional subspace,
\begin{equation}                                                           
\epsilon_{ABCD} = \frac{\epsilon^{\lambda \mu \nu \rho} f_{\lambda A} f_{\mu B} f_{\nu C} f_{\rho D}}{\sqrt{\det \| f_{\lambda A} f^A_\mu \|}}.
\end{equation}

\noindent It is a coordinate function. The $S_{SO(10)}$ can be equivalently rewritten like (\ref{Cartan}) as
\begin{equation}                                                           
S_{SO(10)} = \int \epsilon^{\lambda \mu \nu \rho} \epsilon_{ABCD} f^A_\lambda f^B_\mu R_{\nu \rho}{}^{CD} (\omega ) \d^4 x.
\end{equation}

\noindent The $\epsilon_{ABCD}$ becomes a function of the 4-simplex on the piecewise constant simplicial ansatz. In edge components
\begin{equation}                                                           
\epsilon_{ABCD} (\sigma^4 ) = \frac{\epsilon^{\tson_1 \tson_2 \tson_3 \tson_4} f_{\tson_1 A} f_{\tson_2 B} f_{\tson_3 C} f_{\tson_4 D}}{\sqrt{\det \| f_{\son_1 A} f^A_{\son_2} \|}}.
\end{equation}

\noindent Here $\epsilon^{\tson_1 \tson_2 \tson_3 \tson_4} = \pm 1$ is parity of permutation $(\tson_1 \tson_2 \tson_3 \tson_4 )$ of a quadruple of edges $(\son_1 \son_2 \son_3 \son_4 )$ which span the given 4-simplex. The sum over permutations is implied.

Let an action be a function of edge vectors and of the set of SO(4) rotations as additional variables (connections). It is a discrete analog of the Cartan-Weyl one (\ref{Cartan}). The additional SO(4) variables can be viewed as rotations attributed to each 3-simplex and connecting the frames of the two 4-simplices sharing this 3-simplex. Such action which gives exactly Einstein action on the piecewise flat manifold (Regge action \cite{Regge}) when these additional variables are excluded with the help of eqs of motion was considered in our paper ref. \cite{our2}. We have used the discrete analogs of the connection and curvature first considered in ref. \cite{Fro}. Above consideration allows to transfer the result of ref. \cite{our2} to the considered case of the 4-dimensional manifold and SO(10) rotations. This action has the form
\begin{equation}\label{S discr}                                            
S^{\rm discr}_{SO(10)} = \sum_{\sigma^2} A(\sigma^2 ) \alpha (\sigma^2 ), ~~~ \alpha (\sigma^2 ) = \arcsin \left [ \frac{1}{4} \epsilon_{ABCD} (\sigma^4 ) \frac{V^{AB}_{\sigma^2}}{A(\sigma^2 )} R^{CD}_{\sigma^2} ( \Omega ) \right ].
\end{equation}

\noindent Here
\begin{equation}                                                           
V^{AB}_{\sigma^2} = \frac{1}{2} ( f^A_{\sigma^1_1} f^B_{\sigma^1_2} - f^B_{\sigma^1_1} f^A_{\sigma^1_2} )
\end{equation}

\noindent is bivector of the triangle $\sigma^2$ built on a double of its edge vectors $f^A_{\sigma^1_1}$, $f^A_{\sigma^1_2}$. The area of this triangle
\begin{equation}                                                           
A(\sigma^2 ) = \sqrt{\frac{1}{2} V^{AB}_{\sigma^2} V_{AB \sigma^2} }.
\end{equation}

\noindent The {\it curvature} SO(10) matrix $R^{AB}_{\sigma^2} ( \Omega )$ on the triangles $\sigma^2$ is holonomy of the {\it connection} SO(10) matrix $\Omega^{AB}_{\sigma^3}$ on the 3-simplices (tetrahedrons) $\sigma^3$. That is, $R_{\sigma^2}$ is the product of $\Omega_{\sigma^3}$'s for the set of $\sigma^3$'s containing $\sigma^2$ ordered along the path which encircles $\sigma^2$,
\begin{equation}                                                           
R_{\sigma^2} = \prod_{ \{ \sigma^3 : ~ \sigma^3\supset\sigma^2 \} }{\Omega^{\pm 1}_{\sigma^3}}.
\end{equation}

\noindent This path begins and ends in a 4-simplex $\sigma^4$. That is, $R^{AB}_{\sigma^2}$ is defined in (the frame of) this simplex. The $V^{AB}_{\sigma^2}$ (as well as $f^A_{\sigma^1_1}$, $f^A_{\sigma^1_2}$ constituting this bivector) is also defined in this simplex, and the same simplex appears as argument in $\epsilon_{ABCD} (\sigma^4 )$. Besides pointing out this $\sigma^4$ as argument, we shall also provide notation for this 4-simplex separated by the vertical line, "$| \sigma^4$", in the subscript on the considered value, e. g. $V^{AB}_{\sigma^2}{}_{ | \sigma^4}$. The dual bivector is
\begin{equation}                                                           
v_{\sigma^2 AB | \sigma^4} = \frac{1}{2} \epsilon_{ABCD} (\sigma^4 ) V^{CD}_{\sigma^2}{}_{ | \sigma^4}.
\end{equation}

\noindent The considered $\sigma^4$ in which geometrical values are defined depends on the 2-simplex $\sigma^2$ whose contribution to action is evaluated, i. e. it is function of $\sigma^2$: $\sigma^4 = \sigma^4 (\sigma^2 )$. Thus eq. (\ref{S discr}) can be rewritten as
\begin{equation}                                                           
S^{\rm discr}_{SO(10)} = \sum_{\sigma^2} A(\sigma^2 ) \alpha (\sigma^2 ), ~~~ \alpha (\sigma^2 ) = \arcsin \left [ \frac{v_{\sigma^2 AB | \sigma^4 (\sigma^2 ) }}{2 A(\sigma^2 )} R^{AB}_{\sigma^2}{}_{ | \sigma^4 (\sigma^2 ) } ( \Omega ) \right ].
\end{equation}

To write out the equations of motion for $\Omega_{\sigma^3}$, we add to action the orthogonality condition for $\Omega_{\sigma^3}$ multiplied by the Lagrange multiplier,
\begin{equation}                                                           
S^{\rm discr}_{SO(10)} = \sum_{\sigma^2} A(\sigma^2 ) \arcsin \left [ \frac{v_{\sigma^2 AB | \sigma^4 (\sigma^2 ) }}{2 A(\sigma^2 )} R^{AB}_{\sigma^2}{}_{ | \sigma^4 (\sigma^2 ) } ( \Omega ) \right ] + \sum_{\sigma^3} \mu_{\sigma^3 AB} (\Omega^{CA}_{\sigma^3} \Omega_{\sigma^3 C}{}^B - \delta^{AB} ).
\end{equation}

\noindent Here $\mu_{\sigma^3 AB} = \mu_{\sigma^3 BA}$. The $R_{\sigma^2}$ depends on $\Omega_{\sigma^3}$ only if $\sigma^2 \subset \sigma^3$, and in this case $R_{\sigma^2}$ can be written either as $\Gamma_1 (\sigma^2 , \sigma^3 ) \Omega_{\sigma^3} \Gamma_2 (\sigma^2 , \sigma^3 )$ or as $[ \Gamma_1 (\sigma^2 , \sigma^3 ) \Omega_{\sigma^3} \Gamma_2 (\sigma^2 , \sigma^3 ) ]^{\rm T}$. Consider the former possibility (the latter one differs by the sign of the contribution to $S^{\rm discr}_{SO(10)}$), $R = \Gamma_1  \Omega \Gamma_2 $. Here $\Gamma_1 (\sigma^2 , \sigma^3 )$, $\Gamma_2 (\sigma^2 , \sigma^3 )$ are some SO(10) matrices, the products of $\Omega$'s different from the given $\Omega_{\sigma^3}$. Acting on $S^{\rm discr}_{SO(10)}$ by the operator $( \Omega_{\sigma^3 C}{}^A \partial / \partial \Omega_{\sigma^3 CB} - \Omega_{\sigma^3 C}{}^B \partial / \partial \Omega_{\sigma^3 CA} )$ we cancel the $\mu_{\sigma^3}$-part and get eqs. of motion
\begin{equation}\label{dS/dOmega}                                          
\sum_{\{\sigma^2 : ~ \sigma^2 \subset \sigma^3 \} } \Gamma_2 (\sigma^2 , \sigma^3 ) \frac{v_{\sigma^2 | \sigma^4 (\sigma^2 )} R_{\sigma^2 | \sigma^4 (\sigma^2 )} + R^{\rm T}_{\sigma^2 | \sigma^4 (\sigma^2 )} v_{\sigma^2 | \sigma^4 (\sigma^2 )}}{\cos \alpha (\sigma^2 )} \Gamma^{\rm T}_2 (\sigma^2 , \sigma^3 ) = 0.
\end{equation}

\noindent Let us suppose for a moment that $f^A_{\sigma^1 | \sigma^4_1} f_{\sigma^1 A | \sigma^4_1} = f^A_{\sigma^1 | \sigma^4_2} f_{\sigma^1 A | \sigma^4_2}$ $\forall \sigma^4_1 \supset \sigma^1$ and $\forall \sigma^4_2 \supset \sigma^1$. That is, we can define edge length $l^2_{\sigma^1} = f^A_{\sigma^1} f_{\sigma^1 A}$ independently of the 4-simplex to which the edge $\sigma^1$ belongs. (In other words, transverse components of the piecewise flat metric $g_{\lambda \mu}$ are continuous.) Let us perform an SO(10) rotation for each 4-simplex $\sigma^4$ (the gauge rotation) which forces a quadruple of independent vectors of edges $f^A_{\sigma^1_i | \sigma^4}$, $i = 1, 2, 3, 4$ (and therefore any vector of edge of this 4-simplex) have nonzero components only at $A = 1, 2, 3, 4$, i. e. have the sense of the usual tetrad. The piecewise flat manifold with these edge lengths and local frames formed by these quadruples possesses certain defect angles, curvature matrices $R_{\sigma^2}$ from SO(4) (subgroup of S(10)) which rotate around $\sigma^2$ through these angles and SO(4) connection matrices which provide the values of these curvature matrices. A priori $\Omega_{\sigma^3}$ are arbitrary SO(10) matrices, but if these are set to be equal to these connections, eq. (\ref{dS/dOmega}) is fulfilled simply as the closure condition for the 2-dimensional surface of the 3-simplex $\sigma^3$. Then $\alpha (\sigma^2 )$ is angle defect at the 2-simplex $\sigma^2$ and $S^{\rm discr}_{SO(10)}$ is Regge action.

Next, imposing no condition on $f^A_{\sigma^1 | \sigma^4}$, let us take into account the discrete analog of the condition on $\omega_{\lambda AB}$ (\ref{wff}). Now antisymmetric part of $\Omega_{\sigma^3 AB}$ replaces the continuum connection $\omega_{\lambda AB}$ there\footnote{Some ambiguity lies in the possible choice of antisymmetric part of $\Omega_{\sigma^3 AB}$ mentioned or, say, generator of $\Omega_{\sigma^3 AB}$. Our actual choice of antisymmetric part of $\Omega_{\sigma^3 AB}$ is singled out by the simplest functional dependence on $\Omega$ (linear).}. Therefore this condition takes the form
\begin{equation}\label{Omega V}                                            
\Omega_{\sigma^3 AB} V^{AB}_{\sigma^2}{}_{| \sigma^4 (\sigma^3 )} = 0 ~~~ \forall \sigma^2 \subset \sigma^4 (\sigma^3 ).
\end{equation}

\noindent Here $\sigma^4 (\sigma^3 )$ (function of $\sigma^3$) is one of the two 4-simplices sharing $\sigma^3$. It is not difficult to show that condition (\ref{Omega V}) is equivalent to
\begin{equation}\label{Omega v}                                            
\Omega_{\sigma^3 AB} v^{AB}_{\sigma^2}{}_{| \sigma^4 (\sigma^3 )} = 0 ~~~ \forall \sigma^2 \subset \sigma^4 (\sigma^3 )
\end{equation}

\noindent (in fact, the dual bivector $v^{AB}_{\sigma^2}$ for a triangle $\sigma^2 \subset \sigma^4$ is a combination of the bivectors $V^{AB}_{\sigma^2}$ for the set of the 2-simplices $\sigma^2 \subset \sigma^4$).

As a result, the discrete version of the full action (\ref{S full}) takes the form
\begin{eqnarray}                                                           
S^{\rm discr} & = & \sum_{\sigma^2} A(\sigma^2 ) \arcsin \left [ \frac{v_{\sigma^2 AB | \sigma^4 (\sigma^2 ) }}{2 A(\sigma^2 )} R^{AB}_{\sigma^2}{}_{ | \sigma^4 (\sigma^2 ) } ( \Omega ) \right ] + \sum_{\sigma^3} \mu_{\sigma^3 AB} (\Omega^{CA}_{\sigma^3} \Omega_{\sigma^3 C}{}^B - \delta^{AB} ) \nonumber \\ & & + \sum_{\sigma^3} \sum_{\{ \sigma^2 : ~ \sigma^2 \subset \sigma^4 (\sigma^3 ) \} } \Lambda (\sigma^2, \sigma^3 ) \Omega^{AB}_{\sigma^3} v_{\sigma^2 AB | \sigma^4 (\sigma^3)}.
\end{eqnarray}

\noindent Here $\Lambda (\sigma^2, \sigma^3 )$ are the Lagrange multipliers. More accurately, the last sum is performed not over all $\sigma^2 \subset \sigma^4 (\sigma^3 )$, but over six independent bivectors in $\sigma^4 (\sigma^3 )$ for each $\sigma^3$ (that is, $\Lambda (\sigma^2, \sigma^3 )$ for some $\sigma^2$'s can be set equal to zero).

The equations of motion (resulting from $S^{\rm discr}$ via action of $ ( \Omega_{\sigma^3 C}{}^A \partial / \partial \Omega_{\sigma^3 CB} $ $-$ $ \Omega_{\sigma^3 C}{}^B \partial / \partial \Omega_{\sigma^3 CA} )$ ) take the form
\begin{eqnarray}\label{dSfull/dOmega}                                      
\hspace{-10mm} \sum_{\{\sigma^2 : ~ \sigma^2 \subset \sigma^3 \} } \left [ \Gamma_2 (\sigma^2 , \sigma^3 ) \frac{v_{\sigma^2 | \sigma^4 (\sigma^2 )} R_{\sigma^2 | \sigma^4 (\sigma^2 )} + R^{\rm T}_{\sigma^2 | \sigma^4 (\sigma^2 )} v_{\sigma^2 | \sigma^4 (\sigma^2 )}}{\cos \alpha (\sigma^2 )} \Gamma^{\rm T}_2 (\sigma^2 , \sigma^3 ) \right ]^{AB} & & \nonumber \\ + \sum_{\{ \sigma^2 : ~ \sigma^2 \subset \sigma^4 (\sigma^3 ) \} } \Lambda (\sigma^2, \sigma^3 ) [v_{\sigma^2 | \sigma^4 (\sigma^3 )} \Omega_{\sigma^3} + \Omega^{\rm T}_{\sigma^3} v_{\sigma^2 | \sigma^4 (\sigma^3 )}]^{AB} & = & 0.
\end{eqnarray}

\noindent Excluding $\Lambda (\sigma^2, \sigma^3 )$ we get the weakened form of the equations (\ref{dS/dOmega}): the number of the (combinations of) components available is smaller by six (for the given 3-simplex $\sigma^3$). Instead, there are six newly introduced equations ${\rm Tr} \, \Omega_{\sigma^3} v_{\sigma^2} = 0$.

\section{Correspondence between the Faddeev action on the piecewise constant fields and the discrete first order formalism}

Consider the order of magnitude and contribution of $\Lambda (\sigma^2, \sigma^3 )$ w. r. t. the above introduced typical variation $\delta f$ of $f^\lambda_A$ when passing from simplex to simplex. It is assumed that $\delta f$ is much smaller than the values of the components of $f$ themselves. The continuum connection is linear in the derivatives. This means that the discrete analog $\Omega$ differs from 1 by $O(\delta f)$. If $\Omega = 1$, the first sum in the equations (\ref{dSfull/dOmega}) is (algebraic) sum of the bivectors for closed surface of the 3-simplex. Up to $O(\delta f)$, these bivectors can be assumed to be defined in the same $\sigma^4$. Therefore the sum of these bivectors is zero,
\begin{equation}\label{sum v}                                              
\sum_{\{ \sigma^2 : ~ \sigma^2 \subset \sigma^3 \}} \pm v^{AB}_{\sigma^2} = 0
\end{equation}

\noindent (the closure condition). This is due to vanishing algebraic sum of edge vectors of any triangle by construction (equation (\ref{Sum f})). Therefore the first sum in (\ref{dSfull/dOmega}) is $O(\delta f)$. Let us project these equations horizontally over both indices $A, B$. This amounts to evaluating trace with the set of (six) independent bivectors $v_{\tilde{\sigma}^2 | \sigma^4 (\sigma^3 )}$, i. e. to the action of the operator ${\rm Tr} \, v_{\tilde{\sigma}^2 | \sigma^4 (\sigma^3 )} (\cdot )$ on both sides of the equations (\ref{dSfull/dOmega}). The resulting equations can be solved in regular way for (six per $\sigma^3$) unknowns $\Lambda (\sigma^2, \sigma^3 )$ which thus have an order of magnitude $O(\delta f)$. Having defined $\Lambda (\sigma^2, \sigma^3 )$, we can project the equations (\ref{dSfull/dOmega}) over one of the indices $A, B$ vertically, i. e. apply the projector $\Pi_{CA} \equiv \Pi_{CA} (\sigma^4 (\sigma^3 ))$. Then the second sum becomes combination of expressions vanishing at $\Omega = 1$ with the coefficients $\Lambda (\sigma^2, \sigma^3 )$. That is, this sum is combination of expressions $O(\delta f)$ with the coefficients $\Lambda (\sigma^2, \sigma^3 )$. Thus, this sum is $O((\delta f)^2)$. Therefore, if the order $O(\delta f)$ is considered, the $\Lambda$-part can be omitted in the equations of motion for $\Omega$ projected in the vertical direction.

Taking into account these considerations, let us find solution of the equations (\ref{dSfull/dOmega}) for $\omega_{\sigma^3} = - \omega_{\sigma^3}^{\rm T}$, $\Omega_{\sigma^3} = \exp \omega_{\sigma^3}$, in the order $O(\delta f)$. More precisely, we take natural discretization of the result for the continuum $\omega_{\lambda AB}$ and check that it satisfies the considered discrete equations.

Namely, in the continuum theory the expression for the covariant (w. r. t. the local SO(10)) derivative ${\cal D}_\lambda X_A \d x^\lambda = X_A (x + \d x ) + \omega_{\lambda AB} \d x^\lambda X_B - X_A (x)$ means that the matrix $\delta_{AB} + \omega_{\lambda AB} \d x^\lambda$ transforms (via the parallel transport) $X_A (x + \d x )$ to a vector defined at the point $x$. At the same time, the expression for $\omega_\lambda$ can be written as \cite{our}
\begin{equation}\label{omega cont}                                         
\omega_{\lambda AB} \d x^\lambda f^{\mu B} = \Pi_{AB} [f^{\mu B} (x ) - f^{\mu B} (x + \d x ) ].
\end{equation}

\noindent And, at the same time, the contribution to the Lagrangian density from the curvature developed, say, in the plane $x^1, x^2$ has the form
\begin{equation}                                                           
\sqrt{g} f^1_A f^2_B R^{AB}_{12}, ~~~ R_{12} = [\partial_1 + \omega_1, \partial_2 + \omega_2 ].
\end{equation}

\noindent The $R^{AB}_{12}$ is infinitesimal rotation of a vector transported in parallel way along the closed path successively forward along the coordinate $x^2$, forward along $x^1$, backward along $x^2$ and backward along $x^1$. That is, the path along which a vector is transported, is traversed clockwise in the plane $x^1, x^2$ (for the usual orientation of the coordinate axes in the plane when the shortest rotation of $x^1$ to $x^2$ is counterclockwise).

Now rewrite these continuum result and sign conventions for the discrete case. The discrete analog of $\omega_\lambda \d x^\lambda$ is $\omega_{\sigma^3}$, the generator of $\Omega_{\sigma^3}$. The naive discrete analog of the connection (\ref{omega cont}) is
\begin{equation}\label{omega disc}                                         
\omega_{\sigma^3_j AB} f^{\mu B} (\sigma^4_j ) = \Pi_{AB} (\sigma^4_j ) [f^{\mu B} (\sigma^4_j ) - f^{\mu B} (\sigma^4_{j+1} ) ] + O((\delta f)^2 )
\end{equation}

\noindent for typical neighborhood of a 2-simplex $\sigma^2$ of fig. \ref{sigma2}. It is possible that $\Pi_{AB} (\sigma^4_{j+1} )$ stands for $\Pi_{AB} (\sigma^4_j )$ here, depending on the specific choice ($\sigma^4_j$ or $\sigma^4_{j+1}$) for $\sigma^4 (\sigma^3_j)$ in the additional condition (\ref{Omega v}), see below. The difference between the RHSs of the equation (\ref{omega disc}) for these two choices is $O((\delta f)^2 )$. The contribution of the 2-simplex $\sigma^2$ of fig. \ref{sigma2} to $S^{\rm discr}$ is
\begin{equation}\label{ffR}                                                
A \arcsin \left [ \frac{V}{4 A} (f^1_A f^2_B - f^2_A f^1_B ) ( \Omega_1 \dots \Omega_i \dots \Omega_n )^{AB} \right ].
\end{equation}

\noindent Here $V = \sqrt{g}, A = \sqrt{V^{AB} V_{AB} / 2} = \sqrt{v^{AB} v_{AB} / 2}, V^{AB} = (f^A_3 f^B_4 - F^A_4 f^B_3) / 2, v_{AB} = (f^1_A f^2_B - f^2_A f^1_B )V /2$, and all these values are taken in $\sigma^4_1$ (see fig. \ref{sigma2}). The $f^A_3, f^A_4$ are the vectors of a double of edges 3, 4 of the triangle $\sigma^2$. (Though, the calculation of interest is done in a more direct way just in the variables $f^1_A, f^2_A$.) The $\Omega_j \equiv \Omega_{\sigma^3_j} = \exp \omega_{\sigma^3_j} \equiv \exp \omega_j$ transforms a vector in $\sigma^4_{j+1}$ to the vector in $\sigma^4_j$. Besides that, the curvature matrix $\Omega_1 \dots \Omega_n$ in equation (\ref{ffR}) is transformation of a vector after parallel transport {\it clockwise} around $\sigma^2$ in fig. \ref{sigma2}. Finally, let us write the additional condition (\ref{Omega v}) in the following notation,
\begin{equation}\label{Omega ff}                                           
(\Omega^{AB}_i - \Omega^{BA}_i) f^\lambda_A (\sigma^4_{\#}) f^\mu_B (\sigma^4_{\#}) = 0.
\end{equation}

\noindent Here $\sigma^4_{\#}$ is $\sigma^4_i$ or $\sigma^4_{i+1}$. Once this choice of $\sigma^4_{\#}$ is made, it is the same for each $\lambda$, $\mu$ at the given $i$.

To write out the equations of motion for $\Omega_i$, it is convenient to expand relevant ($\Omega_i$-dependent) contributions to the action of the type of (\ref{ffR}) up to $\omega^2$-terms. Then we can vary these w. r. t. $\omega_i$ and project vertically over one of the SO(10) indices. According to the above said (in the two paragraphs with the equations (\ref{dSfull/dOmega}), (\ref{sum v}) ) this gives the equations of motion in the order $O(\omega ) = O(\delta f )$. Then we can check validity of these equations on the discrete version for connection (\ref{omega disc}). The part of the expression (\ref{ffR}) of interest (the terms $O(\omega^3)$ are omitted) is
\begin{equation}\label{ffR approx}                                         
V (f^{1A} f^{2B} - f^{2A} f^{1B} ) \left [ \omega_i + \omega_i \sum^n_{j = i + 1} \omega_j + \left ( \sum^{i - 1}_{j = 1} \omega_j \right ) \omega_i \right ]_{AB}.
\end{equation}

\noindent Occurring in this expression the sum is equal to
\begin{eqnarray}                                                           
f^{\lambda A} \sum^{i - 1}_{j = 1} \omega_{j AC} & = & - \sum^{i - 1}_{j = 1} \omega_{j CA} f^{\lambda A} (\sigma^4_j) + O((\delta f )^2) \nonumber \\ & = & - \sum^{i - 1}_{j = 1} \Pi_{CA} (\sigma^4_j) [f^{\lambda A} (\sigma^4_j) - f^{\lambda A} (\sigma^4_{j + 1})] + O((\delta f )^2) \nonumber \\ & = & - \Pi_{CA} (\sigma^4_\#) \sum^{i - 1}_{j = 1} [f^{\lambda A} (\sigma^4_j) - f^{\lambda A} (\sigma^4_{j + 1})] + O((\delta f )^2) \nonumber \\ & = & \Pi_{CA} (\sigma^4_\#) [f^{\lambda A} (\sigma^4_i) - f^{\lambda A} (\sigma^4_1)] + O((\delta f )^2)
\end{eqnarray}

\noindent and analogously
\begin{equation}                                                           
f^{\lambda B} \sum^n_{j = i + 1} \omega_{j DB} = \Pi_{DB} (\sigma^4_\#) [f^{\lambda B} (\sigma^4_{i+1}) - f^{\lambda B} (\sigma^4_1)] + O((\delta f )^2).
\end{equation}

\noindent Taking into account these equalities and acting by $V^{-1} (\partial / \partial \omega^{AB}_i - \partial / \partial \omega^{BA}_i )$ on the expression (\ref{ffR approx}) we get
\begin{eqnarray}\label{ff}                                                 
& & 2f^1_A (\sigma^4_1) f^2_B (\sigma^4_1) - 2f^2_A (\sigma^4_1) f^1_B (\sigma^4_1) \nonumber \\ & & + \Pi_{AC} (\sigma^4_\#) [f^{1C} (\sigma^4_i) + f^{1C} (\sigma^4_{i+1}) - 2 f^{1C} (\sigma^4_1 )] f^2_B (\sigma^4_1) \nonumber \\ & & - \Pi_{AC} (\sigma^4_\#) [f^{2C} (\sigma^4_i) + f^{2C} (\sigma^4_{i+1}) - 2 f^{2C} (\sigma^4_1 )] f^1_B (\sigma^4_1) \nonumber \\ & & + f^1_A (\sigma^4_1) \Pi_{BC} (\sigma^4_\#) [f^{2C} (\sigma^4_i) + f^{2C} (\sigma^4_{i+1}) - 2 f^{2C} (\sigma^4_1 )] \nonumber \\ & & - f^2_A (\sigma^4_1) \Pi_{BC} (\sigma^4_\#) [f^{1C} (\sigma^4_i) + f^{1C} (\sigma^4_{i+1}) - 2 f^{1C} (\sigma^4_1 )].
\end{eqnarray}

\noindent Let us add the expression
\begin{equation}                                                           
2 \Pi_{AC} (\sigma^4_\#) \Pi_{BD} (\sigma^4_\#) [f^{1C} (\sigma^4_1) f^{2D} (\sigma^4_1) - f^{2C} (\sigma^4_1) f^{1D} (\sigma^4_1)]
\end{equation}

\noindent to the result (\ref{ff}). Since $\Pi_{AB} (\sigma^4_\#) f^{\lambda B} (\sigma^4_1) = \Pi_{AB} (\sigma^4_\#) [ f^{\lambda B} (\sigma^4_1) - f^{\lambda B} (\sigma^4_\#) ] = O (\delta f)$, this addend is $O((\delta f)^2)$. At the same time, this allows to rewrite the quadratic in $f^\lambda_A (\sigma^4_1 )$ terms concisely using the horizontal projector $\Pi_{\| AB} \equiv \delta_{AB} - \Pi_{AB}$. Besides that, let us introduce the notation $\sigma^4_*$ for the 4-simplex which is complementary to $\sigma^4_\#$ in the set $\{\sigma^4_i, \sigma^4_{i+1}\}$. That is,
\begin{eqnarray}                                                           
\sigma^4_* = \sigma^4_{i+1} ~~~ {\rm if} ~~~ \sigma^4_\# = \sigma^4_i, \nonumber \\ \sigma^4_* = \sigma^4_i ~~~ {\rm if} ~~~ \sigma^4_\# = \sigma^4_{i+1}.
\end{eqnarray}

\noindent Then $\Pi_{AB} (\sigma^4_\# ) [f^{\lambda B} (\sigma^4_i ) + f^{\lambda B} (\sigma^4_{i + 1} )]$ = $\Pi_{AB} (\sigma^4_\# ) f^{\lambda B} (\sigma^4_* )$ = $\Pi_{AB} (\sigma^4_\# ) [ f^{\lambda B} (\sigma^4_* ) - f^{\lambda B} (\sigma^4_\# )]$ = $O(\delta f)$. These values are multiplied in the expression (\ref{ff}) by the linear in $f^\lambda_A (\sigma^4_1 )$ factors. Therefore if we replace $f^\lambda_A (\sigma^4_1 )$ by $f^\lambda_A (\sigma^4_* )$ in the linear over $f^\lambda_A (\sigma^4_1 )$ part of the expression (\ref{ff}), this means variation of this expression by $O((\delta f)^2)$. Thus, the equation (\ref{ff}) takes the form
\begin{eqnarray}                                                           
& & 2 \Pi_{\| AC} (\sigma^4_\#) \Pi_{\| BD} (\sigma^4_\#) [f^{1C} (\sigma^4_1) f^{2D} (\sigma^4_1) - f^{2C} (\sigma^4_1) f^{1D} (\sigma^4_1)] \nonumber \\ & & + \Pi_{AC} (\sigma^4_\#) [ f^{1C} (\sigma^4_* ) f^2_B (\sigma^4_* ) - f^{2C} (\sigma^4_* ) f^1_B (\sigma^4_* ) ] \nonumber \\ & & + \Pi_{BC} (\sigma^4_\#) [ f^1_A (\sigma^4_* ) f^{2C} (\sigma^4_* ) - f^2_A (\sigma^4_* ) f^{1C} (\sigma^4_* ) ].
\end{eqnarray}

\noindent Next we project this by $\Pi_{EA} (\sigma^4_\# )$. This kills the first term, converts the value of the third term to $O((\delta f)^2)$ and leaves us with the second term which is of the order of $O(\delta f)$. Also we can restore the volume factor $V = V (\sigma^4_1 )$. Change it to $V (\sigma^4_* )$ (this leads to negligible possible correction $O((\delta f)^2)$). The resulting expression reads
\begin{eqnarray}\label{Pi ff}                                              
\Pi_{EC} (\sigma^4_\#) [f^{1C} (\sigma^4_*) f^{2B} (\sigma^4_*) - f^{2C} (\sigma^4_*) f^{1B} (\sigma^4_*) ] V (\sigma^4_*) \nonumber \\ = \Pi_{EC} (\sigma^4_\#) \epsilon^{CBAD} (\sigma^4_*) f_{3A} (\sigma^4_*) f_{4D} (\sigma^4_*).
\end{eqnarray}

\noindent Here $f_{3[A} f_{4D]}$ is only one of the bivectors of the four triangles $V_{\sigma^2 AD}$, the faces of $\sigma^3_i$. The $\epsilon^{CBAD} (\sigma^4_*)$ and $f_{\lambda A} (\sigma^4_*)$ depend on the 4-simplex $\sigma^4_*$, but it is important that $\sigma^4_*$ is the same in the contributions like equation (\ref{Pi ff}) of the other faces to the equations of motion. The sum of contributions like expression (\ref{Pi ff}) from all four faces of $\sigma^4_i$ is zero since (algebraic) sum of bivectors is zero for closed surface,
\begin{equation}\label{sum V}                                              
\sum_{\{ \sigma^2 : ~ \sigma^2 \subset \sigma^3_i \}} \pm V_{\sigma^2 AB | \sigma^4_*} = 0.
\end{equation}

\noindent This is due to vanishing algebraic sum of edge vectors of any triangle by construction (\ref{Sum f}). Thus, the connection (\ref{omega disc}) indeed solves the equations of motion for connection in the leading order $O(\delta f)$.

Now it remains to substitute the solution (\ref{omega disc}) into equation (\ref{ffR}) expanded up to $\omega^2$-terms. Since we are finding extremum of the sum of linear and bilinear forms of $\omega$, it is sufficient to evaluate the bilinear form on the solution for $\omega$ found. The result of interest is then that found with the reversed sign,
\begin{equation}\label{ffR approx full}                                    
- V (f^{1A} f^{2B} - f^{2A} f^{1B} ) \left ( \sum_{i > j} \omega_j \omega_j \right )_{AB}.
\end{equation}

\noindent Substituting
\begin{equation}                                                           
f^{\lambda A} \omega_{j A}{}^C \omega_{i CB} f^{\mu B} = - \Pi_{AB} [f^{\lambda A} (\sigma^4_j) - f^{\lambda A} (\sigma^4_{j+1})] [f^{\mu B} (\sigma^4_i) - f^{\mu B} (\sigma^4_{i+1})]
\end{equation}

\noindent we get the contribution to $S^{\rm discr}$
\begin{equation}                                                           
V \Pi^{AB} \sum^n_{i=1} [ f^1_A ( \sigma^4_i ) f^2_B ( \sigma^4_{i+1} ) - f^1_A ( \sigma^4_{i+1} ) f^2_B ( \sigma^4_i ) ].
\end{equation}

\noindent Thus, we have reproduced the model-free $O((\delta f)^2)$ part (important for reproducing the continuum limit) of the Faddeev action for the piecewise-constant ansatz for $f^\lambda_A$, see the equation (\ref{d f d f V Pi}).

\section{Discussion}

As an example, consider hypercubic decomposition of spacetime. This can be viewed as a particular case of the simplicial decomposition if we decompose each hypercube into some number of simplices and then set the fields $f^\lambda_A$ in these simplices to be the same inside the hypercube. As considered above, setting the fields to be constant inside hypercubes does not give any restriction on the form of metric in the Faddeev formulation of gravity which can be approximated by collection of hypercubes (fig. \ref{cubes}). Now curvature residues on quadrangles (plaquettes) rather than the triangles, each quadrangle being the pair of triangles giving the same contribution to action of the type of (\ref{ffR}). Now the number of $\Omega$-matrices in the curvature $n = 4$, these matrices transform a vector being transported along the coordinate directions. We denote by $\Omega_\lambda$ the matrix which acts along (the positive direction of) the coordinate $x_\lambda$ (the $\Omega_\lambda$ transforms a vector at smaller $x_\lambda$ to a vector at larger $x_\lambda$). Also introduce the operator $T_\lambda$ which shifts the argument of a function on the hypercubic lattice from any site (vertex) to the neighboring site along the coordinate $x^\lambda$ (forward). The action takes the form
\begin{eqnarray}                                                           
S^{\rm discr} & = & \sum_{\rm sites} \sum_{\lambda, \mu} \frac{\sqrt{ (f^\lambda )^2 (f^\mu )^2 - (f^\lambda f^\mu )^2}}{2 \sqrt{\det \| f^\lambda f^\mu \|}} \arcsin \left\{ \frac{f^\lambda_A f^\mu_B - f^\mu_A f^\lambda_B}{2 \sqrt{ (f^\lambda )^2 (f^\mu )^2 - (f^\lambda f^\mu )^2}} \right. \nonumber \\ & & \left. \cdot \left[ \Omega_\lambda (T^{\rm T}_\lambda \Omega_\mu) (T^{\rm T}_\mu \Omega^{\rm T}_\lambda) \Omega^{\rm T}_\mu \right]^{AB} \right\} + \sum_{\rm sites} \sum_{\lambda, \mu, \nu} \Lambda^\lambda_{[\mu \nu]} \Omega^{AB}_\lambda (f^\mu_A f^\nu_B - f^\nu_A f^\mu_B).
\end{eqnarray}

\noindent It looks like the sum over plaquettes (quadrangles in $x^\lambda, x^\mu$) in Wilson's discrete action in QCD \cite{Wil}. It possesses the following two properties. First, it is not only some discrete approximation to the exact continuum action but it describes, in principle, the actually existing (minisuperspace) gravity system. Second, $f^\lambda_A$ can be freely chosen in each hypercube (site), that is, neighboring hypercubes do not necessarily coincide on their common faces.

To summarize, we have started with the Faddeev action on $f^\lambda_A (x )$ constant in the interior of each 4-simplex of a simplicial complex. The value of it depends on the model of intermediate regularization of discontinuities of $f^\lambda_A (x )$ between the neighboring 4-simplices. At the same time, this model dependence is negligible compared with the main contribution when $f^\lambda_A (x )$ varies arbitrarily slowly from the 4-simplex to 4-simplex. This slow variation can mean, e. g., the regime of approaching the continuum limit, the model-free main contribution being responsible for recovering the true continuum Faddeev action.

Next we have proposed the discrete form of the connection representation of the Faddeev action on a simplicial complex. We have suggested the connection representation of the Faddeev action (the "first order formalism") earlier. It looks like Cartan-Weyl form of Einstein action generalized to SO(10) plus local SO(10) violating condition, which expresses vanishing the horizontal-horizontal components of the (infinitesimal) connection. Requirement for the discrete form is that if SO(10) violating condition is not imposed, the discrete form of interest should be exact representation of the discrete Einstein (Regge) action (that is, it should result in Regge action upon excluding connections via equations of motion). This fixes the discrete representation of interest practically uniquely, up to non-leading terms in the definition of proper analog of infinitesimal connection in the discrete case when the connection is finite. Another requirement for the discrete form to be fixed, which we have tested, is that the discrete first order formalism is consistent with the above discrete second order formalism (that is, genuine Faddeev action on $f^\lambda_A (x )$ which is piecewise constant on simplices). That is, excluding connections via equations of motion, we reproduce in the leading order (when $f^\lambda_A (x )$ varies arbitrarily slowly from the 4-simplex to 4-simplex) the above model-free main contribution in the second order discrete action responsible for the true continuum limit. At the same time, beyond the leading order we get model-free overall answer which thus can serve definition of the Faddeev action on a piecewise constant ansatz on simplices.

A feature of the Faddeev action is that its existence does not require something like the conditions of continuity of (the transverse components of) the metric. (The continuity eventually is restored on classical level on macroscopic scale when the continuum limit is reached.) That is, the values of the fields in the neighboring 4-simplices can be considered independent. This simplifies description of the system and allows to use an ansatz for which the action is a sum over plaquettes analogous to Wilson's discrete action in QCD.

The author thanks I.A. Taimanov who had attracted author's attention to the new formulation of gravity and Ya.V. Bazaikin for valuable discussion on this subject. The author is grateful to I.B. Khriplovich who has provided moral support, A.A. Pomeransky and A.S.Rudenko for discussion at a seminar, stimulating the writing of this article. The present work was supported in part by the Russian Foundation for Basic Research through Grants No. 09-01-00142-a, 11-02-00792-a and Grant \\ 14.740.11.0082 of federal program "personnel of innovational Russia".


\end{document}